\documentclass[runningheads]{llncs}
\usepackage[T1]{fontenc}
\usepackage{graphicx}
\usepackage{hyperref}  
\usepackage{xurl}
\usepackage{pifont}
\usepackage{color}

\begin{document}
\title{Transforming EU Governance: The Digital Integration through EBSI and GLASS}
\titlerunning{EU Governance: The Digital Integration through EBSI and GLASS}
%
\author{Dimitrios Kasimatis\inst{1}\orcidID{0009-0009-2036-426X} \and
William J Buchanan\inst{1}\orcidID{0000-0003-0809-3523} \and
Mwarwan Abubakar\inst{1} \and
Owen Lo\inst{1}\orcidID{0000-0003-0201-6498} \and
Christos Chrysoulas\inst{1}\orcidID{0000-0001-9817-003X} \and
Nikolaos Pitropakis\inst{1}\orcidID{0000-0002-3392-9970} \and
Pavlos Papadopoulos\inst{1}\orcidID{0000-0001-5927-6026} \and
Sarwar Sayeed\inst{1}\orcidID{0000-0002-9164-7672} \and
Marc Sel\inst{2}\orcidID{0000-0003-3444-1560}}
%

    
\authorrunning{D. Kasimatis et al.}
%
\institute{Edinburgh Napier University, Edinburgh, UK
\\ \and
Royal Holloway, University of London, London, UK
\\}
\maketitle              
\begin{abstract}
Traditionally, government systems managed citizen identities through disconnected data systems, using simple identifiers and paper-based processes, limiting digital trust and requiring citizens to request identity verification documents. The digital era offers a shift towards unique digital identifiers for each citizen, enabling a 'citizen wallet' for easier access to personal documents like academic records and licences, with enhanced security through digital signatures. The European Commission's initiative for a digital wallet for every EU citizen aims to improve mobility and integration, leveraging the European Blockchain Services Infrastructure (EBSI) for harmonised citizen integration. This paper discusses how EBSI and the GLASS project can advance governance and streamline access to identity documents.

\end{abstract}
\section{Introduction}
In recent years, there has been an increased demand from citizens, corporations, and governments to access e-government services. This requires trustworthy e-governance solutions that provide citizens and businesses with safe and transparent public services without compromising their security or privacy. The failure to satisfy these necessities can have severe economic and social consequences. Along with security and privacy, creating an infrastructure that users can rely on is a key consideration when establishing e-government and providing a solid linkage between digital identity and socio-economic inclusion \cite{addo2021advancing}.

Coursey \emph{et al.} predict that \emph{e-government will fundamentally transform the relationship between governments and citizens}  \cite{coursey2008models}. This transformation could lead to improved democracy, including citizen participation, increased transparency, improved trust in government, strengthened democracy, and social inclusion. But approaches to e-governance can only succeed with a citizen focus, increased accountability, citizen empowerment, co-production and good governance \cite{draheim2020narratives}. But, Draheim \emph{et al.}  \cite{draheim2020narratives} also outline that there are critical political topics which must run alongside any development of e-governance applications, including those related to the digital divide, anti-corruption, the loss of privacy, social change, and increasing control from government agencies. 



GLASS \footnote{\label{myfootnote}GLASS is funded within the EU Horizon 2020 research and innovation programme under grant agreement 959879} \cite{lo2022glass} provides a user-centric approach in which digital credentials are referenced within digital wallets owned and controlled by individuals. This eliminates the need for a third party to maintain trust. Citizens then have complete control over their identities and associated data. GLASS employs a trusted framework to enable the requester (often defined as the \emph{relying party}) to validate the verifiability of the issuer's verifiable credentials and digital signature. In the EU, these transfers must be compliant with the General Data Protection Regulation (GDPR) \cite{voigt2017eu} and the incorporation of a legal framework for electronic Identification and Authentication and trust Services (eIDAS). As a result, the GLASS model can carry out procedures that are functional across borders and user-friendly to support  governments, citizens, organisations, and public services. The EBSI infrastructure provides a foundation element of this, which can be used to provide trusted identity-checking services for citizens and trusted organisations. This infrastructure can provide a distributed infrastructure that can reduce administrative burdens and thus connect public administrations across all the EU state members. 

This paper investigates EBSI and GLASS and examines their approaches to handling identity, security and privacy. In addition, we investigate the viable integration of GLASS and EBSI models. Overall, it focuses on a distributed, secure, and scalable e-Government architecture that integrates the GLASS \cite{glass_definition} and EBSI (European Blockchain Services Infrastructure) projects \cite{EBSI}. Both of these projects use permissioned blockchains and distributed ledger methods to ensure security and privacy and address some of the weaknesses of current e-government systems and the interoperability between governance departments. GLASS can use IPFS (InterPlanetary File System) to provide improved resilience.

The rest of the paper is structured as follows. Section \ref{related_work} briefly describes the related work. Section \ref{eu_eid} summarises the European approaches for electronic identities. Soon afterwards, Section \ref{eu_blockchain_part} details the European Blockchain Partnership while Section \ref{sec:Background} provides the necessary background information for the GLASS project. This enables Section \ref{glass_ebsi} to explain the feasibility of combining GLASS with EBSI. Last, Section \ref{sec:Conclusion and future work} concludes while giving some pointers for future work.

\section{Related work}
\label{related_work}

Lykidis \emph{et al.} \cite{lykidis2021use} define a wide range of ongoing blockchain-based e-Government applications including for authentication, e-Voting, land property services, e-Delivery Services, human resources management, and government contracting. Most of these integrate some form of identity management, but the approaches taken vary widely. Rathee and Singh \cite{RATHEE2021} thus provide a literature review of blockchain-related identity management. The two main approaches to identity management can be summarised as: 
\begin{itemize}
\item{using one or more appointed trusted authorities providing assurance over the identity-proofing of subjects,}
\item{allowing subjects to create and manage self-asserted identities, referred to as self-sovereign identities (SSI).}
\end{itemize}

In the EU, blockchain is increasingly considered for e-governance, yet challenges remain in aligning identity linkages with Public Key Infrastructure (PKI) and Qualified Digital Certificates. Turkanovic et al. \cite{turkanovic2020signing} suggest an architecture using CEF components like EBSI, eSignature, and eID compliant with eIDAS.

SSI is increasingly employed in blockchain-based applications, diverging from the traditional reliance on appointed trusted authorities evident in PKI. Various projects, such as China's Blockchain Service Network (BSN) \cite{tang2022blockchain}, exemplify this shift. Similarly, Telechain in India, as defined by Singanamalla \emph{et al.} \cite{singanamalla2022telechain}, collaborates with telecommunication providers to meet regulatory requirements. Estonia's X-Road system, incorporating the Guardtime blockchain solution \cite{buldas2013keyless}, integrates state registries \cite{draheim2020blockchains}. Belarus's cryptohub HTP \cite{bazhenova2018digitalization} signifies its digital economy advancement, defining cryptographic operators and cryptocurrency exchanges. Akpan \emph{et al.} \cite{akpan2022governance} found a significant correlation between e-governance and governance in West Africa, emphasising the integration with established institutions.

SSI enables individuals to create and control their identities using a key pair; transactions are signed digitally with a private key and verified with a public key, the former being securely stored in a citizen wallet inaccessible to others.
This system relates closely to verifiable credentials, as it provides the foundation for their use and management.

Globally, initiatives like The Open Identity Exchange (OIX) and Trust over IP Foundation (ToIP) \cite{ToIP} strive to standardise verifiable credentials and wallets. ToIP emphasises decentralised digital identity, issuing guidelines for Hyperledger Aries, Indy, and verifiable credentials \cite{Dizme2020}. The SSI framework encompasses the user (holder), credential issuer, and relying party, with the user managing credential acquisition and transfer. Trust is established via the issuer's digital signature and verification against their trusted public key.

%



%

Appointed trusted authorities in digital applications ensure identity-proofing for natural persons through centralised or decentralised registers and business registers for legal entities. Anand and Brass \cite{AnandNishant2021Rifd} examine eID systems and their governance.







Centralised data storage can shift to distributed models like IPFS, which uses content hashing for file distribution. However, IPFS struggles with content deletion and limited encryption. Politou \emph{et al.} \cite{politou2020delegated} discuss GDPR's Right to be Forgotten (RtbF), highlighting the challenges in IPFS content erasure. They suggest an anonymous protocol for delegated erasure requests restricted to original content providers or their delegates.

\section{Overview of European Electronic Identity Systems}
\label{eu_eid}

The concept of identity, and by extension, electronic identity (e-ID), differs markedly across nations. Globally, notable e-ID implementations include the ICAO electronic passport system \cite{pasupathinathan2008line}, prevalent in most countries, and India's Aadhaar system \cite{khera2017impact}. Each country's political and cultural context influences trust in these systems. Interestingly, some nations lack a national identity scheme. For instance, the UK abandoned its national ID scheme, influenced partly by a blog post \cite{Saunders2006}. 

\subsection{European Union's Approach to Electronic Identity}
\subsubsection{eIDAS Framework}
\label{subsubsec:eidas}
The European Commission's DG Connect initiated the eIDAS (electronic Identification, Authentication and trust Services) regulation, encompassing e-identification and trust services \cite{eidasreg}. This initiative is technically bolstered by ETSI and CEN, focusing on standard development. Implementing Decisions and Regulations specifically target e-identification \cite{eidasregia-ID-4} and trust services \cite{eidasregia-TS-4}, with additional rules covering e-signature products and Trusted Lists.

Delos et al. \cite{seletal2015eidas} detail how EU member states can independently manage their trust ecosystems. A member state may notify the EU Commission of its identity management systems, leading to mutual recognition across the EU. Minimum identity attributes for natural and legal persons are established \cite{eidasregia-ID-2}. Trust Service Providers (TSPs), including Qualified Trust Service Providers (QTSPs), are monitored by national supervisory bodies. These bodies rely on Conformity Assessment Bodies (CABs), usually private firms, for TSP and QTSP evaluation.

CABs must be accredited by a National Accreditation Body (NAB) and publicly declare their conformity assessment scheme. The European Cooperation for Accreditation (EA) endorses an eIDAS accreditation scheme based on ISO/IEC 17065 \cite{ISO17065:2012} and ETSI EN 319 403 \cite{etsien319403}, as outlined in EA Resolution EA 2014 (34) 22 \cite{EA-resolution-2014-34-22}.

eIDAS Article 3 \cite{tsakalakis2017identity} establishes terminology and definitions for electronic signatures, categorising trust services into basic, advanced, and qualified levels. The EU Commission provides a legal, functional, and technical framework for evaluating and validating identity and trust services, including the publication of signed metadata and the List of Trusted Lists (LOTL).

\subsubsection{European Digital Identity Wallet}
The 2021 proposal \cite{EuropeanCommission2021a} recommended updates to eIDAS, mandating EU states to provide a European Digital Identity Wallet (EDIW). This initiative introduces four new Trust Services, enhancing support for distributed ledger technologies and electronic signature applications.

The pilot phase for EDIWs starts in 2023, aiming for EU-wide availability by 2024. These wallets are expected to be GDPR-compliant and integrate with the eIDAS framework. Thales's survey \cite{Thales} reveals varying perceptions and concerns among EU citizens regarding these wallets' privacy, security, and convenience.

\section{European Blockchain Partnership}
\label{eu_blockchain_part}
In 2018, the European Blockchain Partnership (EBP) was formed by 27 EU states, Norway, Liechtenstein, and later Ukraine \cite{queiruga2022self}. This partnership led to the European Blockchain Services Infrastructure (EBSI), targeting use cases like Self-Sovereign Identity and Diploma awards \cite{vasileedis}, Document Traceability, and Trust Data Sharing. The EBSI uses a public permissioned blockchain, with digital credentials stored in citizen-controlled wallets \cite{grech2021blockchain}. Baldacci et al. \cite{baldacci2021advancing} outline EBSI's core principles, including decentralisation, scalability, open specifications, sustainability, and interoperability.

In 2020, initiatives like DIZME, Findy, Lissi, and MeineSichereID collaborated within the Trust over IP Foundation \cite{EBSI2020}, aiming to create a unified European SSI Network. The European Commission has developed various blockchain strategies \cite{fulli2021policy}, including crypto-asset regulation \cite{sandner2022crypto} and market infrastructure development based on distributed ledger technology \cite{priem2022european}. These efforts are integral to the EBP and its investment in EBSI \cite{EBSI}.

\subsection{EBSI and Its Applications}
\subsubsection{The EBSI Project}
Turkanovic et al. \cite{turkanovic2020signing} describe EBSI's integration with EU public services, running on Hyperledger Besu \cite{dalla2021your} and Fabric clients. EBSI's ledger protocol allows for a pluggable design, adaptable to different client needs \cite{EBSI}. The infrastructure supports public reading access, with writing permissions limited to trusted entities.


EBSI's use cases encompass identity verification, academic awarding, social security checks, and document traceability. The European Self-Sovereign Identity Framework (ESSIF) at EBSI's core is GDPR and e-IDAS compliant, enabling secure and legal identity management for EU citizens.

Bittins et al. \cite{bittins2021healthcare} discuss EBSI's potential for secure healthcare data sharing across the EU, leveraging blockchain technology for data provenance and integrating SSI.

\subsection{EBSI Compliance and Security}
EBSI aligns with several regulations, including the NIS Directive \cite{directive2016directive}, the eIDAS Regulation, and GDPR. Its security framework mandates appropriate measures for safeguarding the infrastructure and the applications built upon it. These measures are essential for maintaining the reliability and integrity of the EBSI network and its services.  

\section{The GLASS Project: A Study of Distributed Ledger Technology}\label{sec:Background}

The GLASS project adopts a distributed domain approach. The project is based on Hyperledger Fabric, a platform used for building blockchain applications. In this instance, three sovereign nations collaborate on a single channel to enhance their governance infrastructure. Each nation comprises two departments responsible for the endorsement, validation, and recording of citizen data. These departments sign identity documents using private keys and verify them with public keys on the EBSI ledger.
  

\subsection{Decentralised Distributed Ledger and Chaincode}
Blockchain technology in the GLASS project is characterised as decentralised, resistant to tampering, and distributed across peer-to-peer networks~\cite{Papadopoulos2020}. Coupled with smart contracts, it provides a structure for developing access control mechanisms vital for secure identification, authentication, and user authorisation. The immutable ledger of the blockchain safeguards against integrity issues. Smart contracts enable the implementation of programmable policies for user authentication and authorisation in a distributed setting~\cite{MarkoHolbl}.

In Hyperledger Fabric, a permissioned blockchain, the chaincode links with the ledger and forms the core of the blockchain system~\cite{chaincode}. The chaincode runs on peers within an organisation, managing the creation, querying, and updating of transactions on the shared ledger. It is instrumental in formulating asset definitions for business contracts and managing decentralised applications.

\subsection{Interplanetary File System}
The InterPlanetary File System (IPFS) is a significant element in the GLASS project~\cite{benet2014ipfs}. This protocol allows for the distributed storage and transfer of digital content using a peer-to-peer model and content-based addressing. Similar to BitTorrent, IPFS's distributed nature facilitates file sharing across a network. IPFS can create a permanent and distributed web system, either public or private. It replaces the traditional URL system with cryptographic hashes as addresses. In the IPFS network, data is distributed without a central point, and peers become distributors of content once downloaded. Digital assets like folders, images, and data stores are segmented and stored distributedly, each piece content-addressed using a Content Identifier (CID).

\subsection{Content Identifier and Distributed Hash Table}
A CID in IPFS serves as a unique marker for resources, focusing on content rather than its location, unlike URLs~\cite{Multiformats2020}. CIDs maintain a fixed length and are cryptographic hashes of the content. Any change in data results in a different CID. The Distributed Hash Table (DHT) acts as a decentralised key-value storage system, serving as a distributed database for storing and retrieving information based on node keys~\cite{benet2014ipfs}. Its main function is to link a CID to the content's location in IPFS.

\subsection{Distributed Infrastructure of Trust and Triplets Design}
The GLASS project's trust infrastructure is based around domains specific to each country. This includes mapping digital signers specific to each country and their rights. Departments like the Department of Justice in Germany (DE) country domain have signing keys, which are stored in Hyperledger along with their public keys, identity (ID), and claim rights.

\begin{figure*}
\vspace{-15 pt}
\begin{center}
\includegraphics[width=0.6\linewidth]{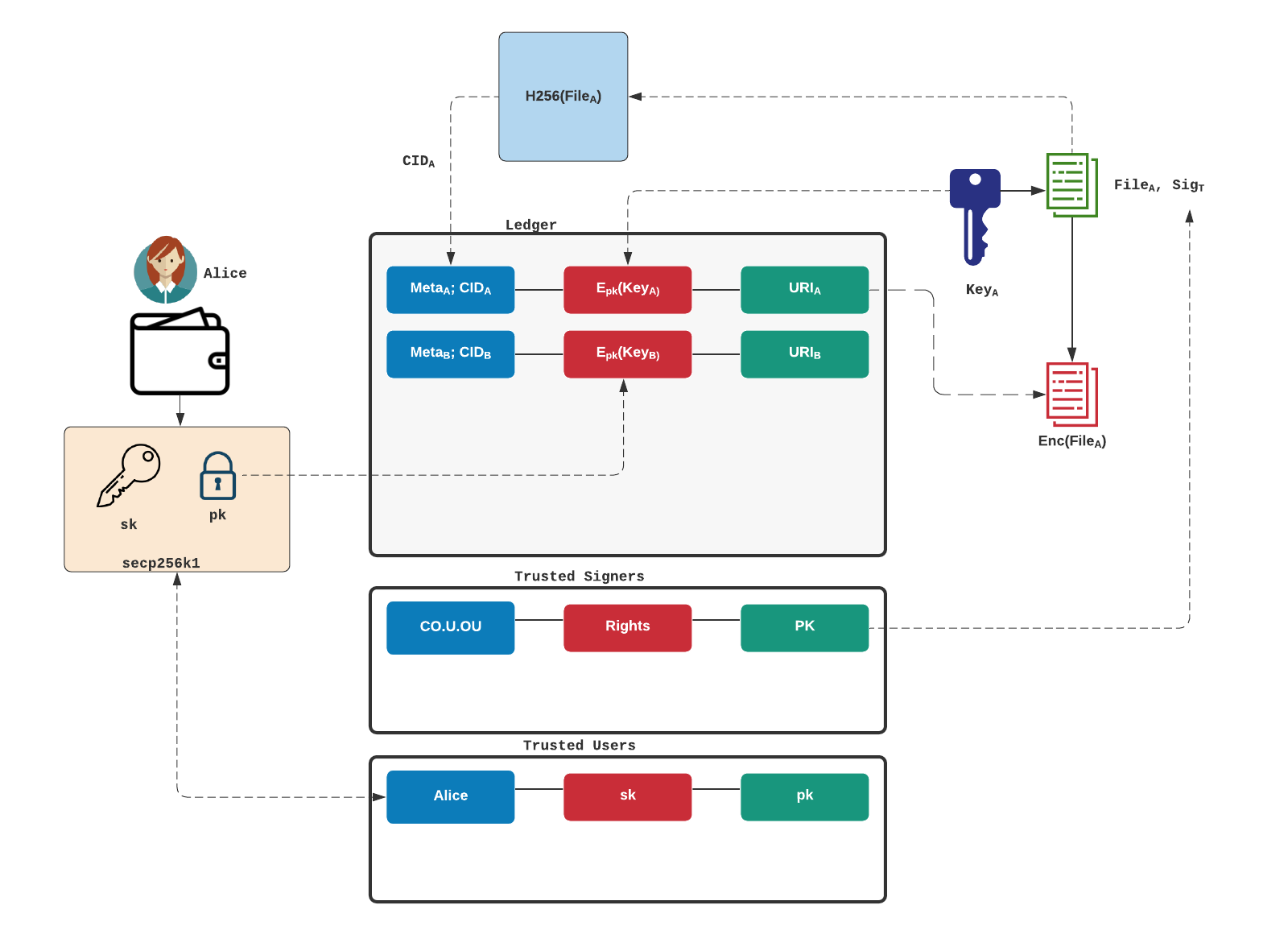}
\caption{Overview of triplets.}
\label{fig:trip01}
\end{center}
\vspace{-20 pt}
\end{figure*}

The triplets design in Hyperledger Fabric is the primary metadata, enabling decryption of data distributed on IPFS. Figure~\ref{fig:trip01} shows each citizen's credential file encrypted with a unique key using the AES GCM method. The encrypted key, along with the file's CID and location (URI), is stored. The URI may link to an IPFS store or a URL.

The trust within each domain is supported by storing the public key of credential signers. These signers are trusted for specific credential types as outlined in the trust policy. The architecture ensures that the URI of the encrypted credential is stored, but the file contents are inaccessible, retrievable only by the citizen's private key.


\subsection{Resource Distribution and Integration}
The GLASS project employs the InterPlanetary File System as its primary protocol for resource distribution. The current prototype encrypts all resources distributed on IPFS, using a private instance of IPFS for enhanced security. This private network operates like the public version, but only nodes with a shared private key can access it, preventing inadvertent exposure of sensitive information.

In integrating Hyperledger Fabric with IPFS, the GLASS project uses both CID and URI to store the same content on IPFS. The separate field property of URI allows future content distribution through various mechanisms, such as Dropbox and SharePoint. In these cases, the CID remains unchanged, but the URI varies based on the resource location. This design allows for the encryption key to be hosted in an external domain, adding flexibility and security to the system.

\section{Analysis of GLASS e-Governance and EBSI Infrastructure Integration}
\label{glass_ebsi}



\subsection{EBSI}
\subsubsection{Business Application}
Integrating GLASS-related business applications within the EBSI framework primarily aims to facilitate the merging of public and private sector applications. This integration occurs either through core service APIs or by hosting a node within the network. The EBSI business interface permits applications to utilise the EBSI-exposed API interfaces, enabling transactions with smart contracts in the EBSI ledger.
\vspace{-15 pt}
\subsubsection{Use Cases}
EBSI use cases to demonstrate its capability to provide public services across international boundaries. These cases are designed to streamline administrative processes and validate the integrity of digital information across various sectors. This approach enhances efficiency and public trust. The initial use cases in EBSI Version 1 include notarisation of documents, diploma management, European Self-Sovereign Identity, and trusted data exchange. Additional use cases, such as SME finance, asylum, and the European Social Security Platform, were subsequently introduced.
\vspace{-15 pt}
\subsubsection{Core Services}
The core services layer includes functionalities applicable to any use case. It provides interfaces for various EBSI services, encompassing both on-chain and off-chain components, and contains application enablers. These microservices are categorised into five distinct groups, offering tools like the Integration API, trusted registries, API security management, and the wallet library. Additionally, digital identity-related features are accessible through this layer.
\vspace{-15 pt}
\subsubsection{Chain and Storage}
This layer comprises supported distributed ledger and storage protocols. The EBSI V2 Ledger API facilitates user interaction with Hyperledger Besu (Read and Write) and Fabric ledgers (Read Only) \cite{li2020survey}. Smart contracts within this layer manage and execute trusted transactions for recording on the blockchain.
\vspace{-15 pt}
\subsubsection{EBSI Infrastructure}
The EBSI infrastructure layer provides general capabilities and connectivity to blockchain networks, including network, compute, and deployment capabilities. It encompasses all components necessary for establishing an EBSI node. The ledger validates, approves, and stores transactions on the blockchain. Presently, 25 nodes hosted by participating member states form this decentralised network.

\subsection{GLASS}
The GLASS architecture, reliant on a trust infrastructure spanning country-wide domains, facilitates distributed trust infrastructure amongst EU member states. Parallel to EBSI, GLASS provides an e-governance framework for use by EU member states, leveraging distributed ledger technologies for efficient, transparent, and user-friendly public service solutions.

\subsubsection{Business Application}
This layer aims to bridge connections between entities legally permitted to utilise GLASS services. It includes a user’s wallet, an essential management tool providing a graphical interface for managing digital keys. These keys assist in signing transactions, credentials, and documents, enabling secure interactions and relationships with third parties.
\vspace{-15 pt}
\subsubsection{Use Cases}
GLASS’s primary use cases involve relocation to another Member State, short-term visits to other countries, and securing employment abroad \cite{glassusecase}.
\vspace{-25 pt}
\subsubsection{Core Services}
The core services of GLASS establish connections between the GLASS portal, Hyperledger Fabric ledger, and a private IPFS network instance, forming a trusted e-governance data-sharing model. These services comprise various identity trust capabilities and APIs, enabling blockchain-based services to achieve high identity assurance and trust standards. Services include credential signing and sharing, ledger transactions, encryption, and APIs for issuing verifiable credentials linked to electronic identification (eID).
\vspace{-15 pt}
\subsubsection{GLASS Portal}
The GLASS portal integrates distributed ledger technologies with a private IPFS network instance for recording and storing encrypted verified credentials on a Hyperledger fabric ledger with restricted access. It facilitates sharing encrypted resources over a private IPFS cluster, creating a Content Identifier (CID) for the resource. The portal also interacts with Hyperledger Fabric to record the CID, URI, and encryption keys. Upon completing these processes, users receive a CID and URI for their encrypted resource, enabling future retrieval.
\vspace{-15 pt}
\subsubsection{Chain and Storage}
GLASS employs IPFS for storage, integrating it with Hyperledger Fabric. The system’s architecture utilises URI, CID, and a Protected Encryption Key for security, which can store these elements in a secure Hyperledger Fabric record within the EBSI architecture. Hyperledger Fabric’s internal ‘chaincode’ facilitates transactions and ledger updates. A specific GLASS-ipfs chaincode in Hyperledger Fabric allows for the creation and reading of Glass resource keys.
\vspace{-15 pt}
\subsubsection{GLASS Ledger}
The GLASS ledger, a permissioned blockchain on Hyperledger Fabric, stores encrypted and disseminated resource triplets (CID, Key, URI) on IPFS. These triplets are stored in hashed form for security. Two types of data collection are utilised: one public (readable by both participating organisations) and one private (accessible only by org1.org). This setup aligns with the proposed Triplet concept.

\subsection{GLASS over EBSI Model}
The GLASS over EBSI model integrates numerous identity trust components for high assurance in blockchain-based services. This model can utilise EBSI services, incorporating eSignature and eID properties for EU cross-border identification. It ensures robust linkages between a user’s digital and physical identities by encrypting and storing credentials on IPFS, verifiable by all EU state members.

\begin{figure*}
\vspace{-22 pt}
\begin{center}
\includegraphics[width=0.6\linewidth]{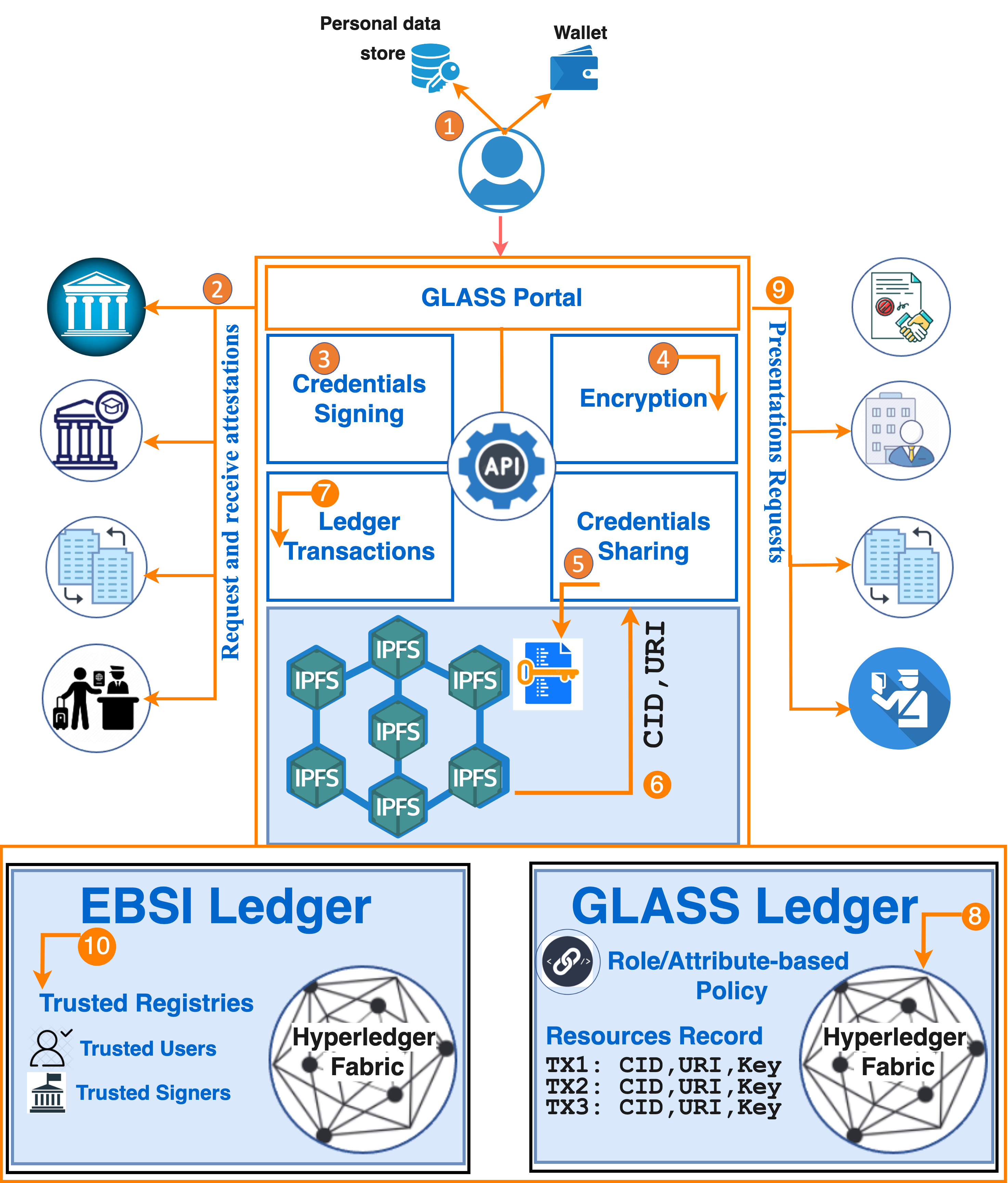}
\caption{GLASS over EBSI model}
\label{fig:overeb}
\end{center}
\vspace{-22 pt}
\end{figure*}

In this model, GLASS provides wallet functionality for citizens and organisations, enabling user identification and association with DIDs in the EBSI ledger. The model is illustrated in Figure \ref{fig:overeb}, depicting identifiers for citizens and organisations (legal and natural persons).

Client wallets use the EBSI wallet API to generate key pairs, initiating the onboarding process for recording information in the EBSI ledger. Authorised issuers and individuals are registered in the EBSI ledger via dedicated APIs, facilitating verification processes within the ecosystem.

The model (Figure \ref{fig:overeb}) outlines various processes:

\begin{itemize}
    \item Students can request and present verifiable university credentials to prospective employers in other countries.
    \item Universities issue encrypted certificates, with GLASS portals facilitating secure resource distribution via private IPFS networks.
    \item The use of AES-256 CBC for encryption and automatic distribution of resources on IPFS is detailed, highlighting security measures.
    \item The GLASS portal records metadata triplets (CID, encryption key, URI) in Hyperledger Fabric data collections for auditing and retrieval.
    \item Verification processes involve students presenting credentials to employers, with verifiers querying the EBSI ledger to confirm issuer DIDs.
\end{itemize}

The GLASS and EBSI integration demonstrates a robust cross-border e-governance and data-sharing framework. This underscores the importance of secure, transparent, and efficient digital identity management in the European Union.

\section{Conclusion}\label{sec:Conclusion and future work}

At its core, enhanced e-governance could increase government services’ efficiency, quality, and effectiveness. However, many political narratives can stall its development, including the digital divide, anti-corruption, the loss of privacy, social change, and the increasing control from government agencies \cite{draheim2020narratives}.  

GLASS \cite{chrysoulas2021glass} focuses on developing an e-governance framework that the European Union’s member states could adopt, while EBSI focuses on providing trusted identity data. Key features of EBSI are the provision of GDPR compliance and eIDAS 2 signatures. The two infrastructures could thus provide an integrated approach to enhanced e-governance services across the EU and support EU citizens' freedom of movement in the future.

\begin{credits}

\subsubsection{\discintname}
The authors have no relevant financial or non-financial interests to disclose.
The authors have no conflicts of interest to declare that are relevant to the content of this article.
All authors certify that they have no affiliations with or involvement in any organisation or entity with any financial interest or non-financial interest in the subject matter or materials discussed in this manuscript.
The authors have no financial or proprietary interests in any material discussed in this article.
\end{credits}
%
%
%
\bibliographystyle{splncs04}
\bibliography{main}

\begin{thebibliography}{10}
\providecommand{\url}[1]{\texttt{#1}}
\providecommand{\urlprefix}{URL }
\providecommand{\doi}[1]{https://doi.org/#1}

\bibitem{addo2021advancing}
Addo, A., Senyo, P.K.: Advancing e-governance for development: Digital identification and its link to socioeconomic inclusion. Government Information Quarterly  \textbf{38}(2),  101568 (2021)

\bibitem{akpan2022governance}
Akpan-Obong, P.I., Trinh, M.P., Ayo, C.K., Oni, A.: E-governance as good governance? evidence from 15 west african countries. Information Technology for Development pp. 1--20 (2022)

\bibitem{AnandNishant2021Rifd}
Anand, N., Brass, I.: Responsible innovation for digital identity systems. Data \& Policy  \textbf{3} (2021)

\bibitem{baldacci2021advancing}
Baldacci, E., Frade, J.R.: Advancing digital transformation in the public sector with blockchain: A view from the european union. In: Disintermediation Economics, pp. 281--295. Springer (2021)

\bibitem{bazhenova2018digitalization}
Bazhenova, T.: Digitalization of belarus economy: prospects of the industry. Section 1 Globalized Economy: Challenges and Prospects p.~21 (2018)

\bibitem{benet2014ipfs}
Benet, J.: Ipfs-content addressed, versioned, p2p file system (draft 3). arXiv preprint arXiv:1407.3561  (2014), \url{https://arxiv.org/abs/1407.3561v1}

\bibitem{bittins2021healthcare}
Bittins, S., Kober, G., Margheri, A., Masi, M., Miladi, A., Sassone, V.: Healthcare data management by using blockchain technology. In: Applications of blockchain in healthcare, pp. 1--27. Springer (2021)

\bibitem{buldas2013keyless}
Buldas, A., Kroonmaa, A., Laanoja, R.: Keyless signatures’ infrastructure: How to build global distributed hash-trees. In: Nordic Conference on Secure IT Systems. pp. 313--320. Springer (2013)

\bibitem{chrysoulas2021glass}
Chrysoulas, C., Thomson, A., Pitropakis, N., Papadopoulos, P., Lo, O., Buchanan, W.J., Domalis, G., Karacapilidis, N., Tsakalidis, D., Tsolis, D.: Glass: Towards secure and decentralized egovernance services using ipfs. In: European Symposium on Research in Computer Security. pp. 40--57. Springer (2021)

\bibitem{coursey2008models}
Coursey, D., Norris, D.F.: Models of e-government: Are they correct? an empirical assessment. Public administration review  \textbf{68}(3),  523--536 (2008)

\bibitem{dalla2021your}
Dalla~Palma, S., Pareschi, R., Zappone, F.: What is your distributed (hyper) ledger? In: 2021 IEEE/ACM 4th International Workshop on Emerging Trends in Software Engineering for Blockchain (WETSEB). pp. 27--33. IEEE (2021)

\bibitem{seletal2015eidas}
Delos, O., Debusschere, T., Soete, M.D., Dumortier, J., Genghini, R., Graux, H., Lacroix, S., Ramunno, G., Sel, M., Eecke, P.V.: A pan-european framework on electronic identification and trust services for electronic transactions in the internal market. In: Paulus, S., Pohlman, N., Reimer, H. (eds.) Securing business processes. pp. 173--195. Vieweg+Tuebner, Springer Science+Business Media (2015)

\bibitem{directive2016directive}
Directive, N.: Directive (eu) 2016/1148 of the european parliament and of the council of 6 july 2016 concerning measures for a high common level of security of network and information systems across the union. OJ L  \textbf{194}(19.7), ~2016 (2016)

\bibitem{Dizme2020}
Dizme: {Position statement toward EBSI} (2022), \url{https://https://lissi.id/ about}

\bibitem{EBSI2020}
Dizme, Lissi, F., MeineSichereID: {Position Statement toward EBSI} (2022), \url{https://networkofnetworks.net/wp-content/uploads/2021/05/EBSI_Network-of-Networks_esatus_Danube_Tech_TNO.pdf}

\bibitem{draheim2020blockchains}
Draheim, D.: Blockchains from an e-governance perspective: Potential and challenges. Proceedings of the EGOSE  (2020)

\bibitem{draheim2020narratives}
Draheim, D., Pappel, I., Lauk, M., Mcbride, K., Misnikov, Y., Nagumo, T., Lemke, F., Hartleb, F.: On the narratives and background narratives of e-government  (2020)

\bibitem{EA-resolution-2014-34-22}
EA: EA Resolution 2014 (34) 22. European cooperation for Accreditation (2014), \url{https://european-accreditation.org/wp-content/uploads/2018/10/34th-ea-ga-approved-resolutions-.pdf}

\bibitem{EBSI}
EBSI: {Architecture - EBSI Documentation -}, \url{https://ec.europa.eu/digital-building-blocks/wikis/display/EBSIDOC/Architecture}

\bibitem{etsien319403}
{ETSI}: Electronic {S}ignatures and {I}nfrastructures ({ESI}) {T}rust {S}ervice {P}rovider {C}onformity {A}ssessment --- {R}equirements for conformity assessment bodies assessing {T}rust {S}ervice {P}roviders (2015), eN 319 403

\bibitem{EuropeanCommission2021a}
{European Commission}: {Proposal for a REGULATION OF THE EUROPEAN PARLIAMENT AND OF THE COUNCIL amending Regulation (EU) No 910/2014 as regards establishing a framework for a European Digital Identity}. Tech. rep. (2021), \url{https://eur-lex.europa.eu/legal-content/EN/TXT/?uri=COM%3A2021%3A281%3AFIN&qid=1622704576563}

\bibitem{eidasreg}
{European Union}: EU 910/2014 Regulation of the European Parliament and of the Council of 23 July 2014 on electronic identification and trust services for electronic transactions in the internal market and repealing Directive 1999/93/EC. DG CONNECT (2014), oJ L 257, 28.8.2014, p. 73 114

\bibitem{eidasregia-ID-2}
{European Union}: EU 2015/1501, Implementing Act on interoperability framework. European Commission (2015), eU 2015/1501

\bibitem{eidasregia-ID-4}
{European Union}: EU 2015/1984, Implementing Act defining the circumstances, formats and procedures of notification. European Commission (2015), eU 2015/1984

\bibitem{eidasregia-TS-4}
{European Union}: EU 2016/650, Standards for the security assessment of qualified signature and seal creation devices. European Commission (2016), eU 2016/650

\bibitem{fulli2021policy}
Fulli, G., Kotzakis, E., Nai~Fovino, I.: Policy and regulatory challenges for the deployment of blockchains in the energy field  (2021)

\bibitem{glass_definition}
{GLASS}: {GLASS for Citizens}. \url{https://www.glass-h2020.eu/for-citizens} (2020), [Online; accessed 10-February-2021]

\bibitem{glassusecase}
GLASS: Moving to another member state (2022), \url{https://www.glass-h2020.eu/moving-to-another-member-state}

\bibitem{grech2021blockchain}
Grech, A., Sood, I., Ari{\~n}o, L.: Blockchain, self-sovereign identity and digital credentials: promise versus praxis in education. Frontiers in Blockchain  \textbf{4},  616779 (2021)

\bibitem{chaincode}
{Hyperledger Fabric}: {Smart Contracts and Chaincode}. \url{https://hyperledger-fabric.readthedocs.io/en/release-2.2/smartcontract/smartcontract.html#system-chaincode} (2020), [Online; accessed 31-Aug-2021]

\bibitem{ISO17065:2012}
ISO: {ISO}/{IEC} 17065:2012(E) {C}onformity assessment --- {R}equirements for bodies certifying products, processes and services

\bibitem{khera2017impact}
Khera, R.: Impact of aadhaar in welfare programmes. Available at SSRN 3045235  (2017)

\bibitem{li2020survey}
Li, D., Wong, W.E., Guo, J.: A survey on blockchain for enterprise using hyperledger fabric and composer. In: 2019 6th International Conference on Dependable Systems and Their Applications (DSA). pp. 71--80. IEEE (2020)

\bibitem{lo2022glass}
Lo, O., Buchanan, W.J., Sayeed, S., Papadopoulos, P., Pitropakis, N., Chrysoulas, C.: Glass: A citizen-centric distributed data-sharing model within an e-governance architecture. Sensors  \textbf{22}(6), ~2291 (2022)

\bibitem{lykidis2021use}
Lykidis, I., Drosatos, G., Rantos, K.: The use of blockchain technology in e-government services. Computers  \textbf{10}(12), ~168 (2021)

\bibitem{MarkoHolbl}
Marko Hölbl Marko~KomparaOrcID, A.K., Zlatolas, L.N.: A systematic review of the use of blockchain in healthcare. MDPI Symmetry  (2018)

\bibitem{Multiformats2020}
multiformats: {multiformats/cid: Self-describing content-addressed identifiers for distributed systems} (2020), \url{https://github.com/multiformats/cid}

\bibitem{Papadopoulos2020}
Papadopoulos, P., Pitropakis, N., Buchanan, W.J.: Decentralised privacy: A distributed ledger approach. In: Hussain, C.M., Di~Sia, P. (eds.) Handbook of Smart Materials, Technologies, and Devices: Applications of Industry 4.0, pp. 1--26. Springer International Publishing, Cham (2020). \doi{10.1007/978-3-030-58675-1_58-1}, \url{https://doi.org/10.1007/978-3-030-58675-1_58-1}

\bibitem{pasupathinathan2008line}
Pasupathinathan, V., Pieprzyk, J., Wang, H.: An on-line secure e-passport protocol. In: International Conference on Information Security Practice and Experience. pp. 14--28. Springer (2008)

\bibitem{politou2020delegated}
Politou, E., Alepis, E., Patsakis, C., Casino, F., Alazab, M.: Delegated content erasure in ipfs. Future Generation Computer Systems  \textbf{112},  956--964 (2020)

\bibitem{priem2022european}
Priem, R.: A european distributed ledger technology pilot regime for market infrastructures: finding a balance between innovation, investor protection and financial stability. Journal of Financial Regulation and Compliance  (2022)

\bibitem{queiruga2022self}
Queiruga-Dios, A., P{\'e}rez, J.J.B., Encinas, L.H.: Self-sovereign identity in university context. In: 2022 31st Conference of Open Innovations Association (FRUCT). pp. 259--264. IEEE (2022)

\bibitem{RATHEE2021}
Rathee, T., Singh, P.: A systematic literature mapping on secure identity management using blockchain technology. Journal of King Saud University - Computer and Information Sciences  (2021). \doi{https://doi.org/10.1016/j.jksuci.2021.03.005}, \url{https://www.sciencedirect.com/science/article/pii/S1319157821000690}

\bibitem{sandner2022crypto}
Sandner, P.G., Ferreira, A., Dunser, T.: Crypto regulation and the case for europe. In: Handbook on Blockchain, pp. 661--693. Springer (2022)

\bibitem{Saunders2006}
Saunders, F.S.: {The Terrible Truth About ID Cards - UK Indymedia} (2006), \url{https://www.indymedia.org.uk/en/2006/04/338263.html}

\bibitem{singanamalla2022telechain}
Singanamalla, S., Mehra, A., Chandran, N., Lohchab, H., Chava, S., Kadayan, A., Bajpai, S., Heimerl, K., Anderson, R., Lokam, S.: Telechain: Bridging telecom policy and blockchain practice. arXiv preprint arXiv:2205.12350  (2022)

\bibitem{tang2022blockchain}
Tang, X., Jia, Z., Yang, W.: Blockchain application status and ecology. In: Blockchain Application Guide, pp. 35--48. Springer (2022)

\bibitem{Thales}
Thales: {Thales Major Citizen Survey Predicts Warm Welcome for New European Digital ID Wallet (EDIW)}, \url{https://www.afp.com/en/news/1312/ \\thales-major-citizen-survey-predicts-warm-welcome-\\new-european-digital-id-wallet-\\ediw-202206050050101, urldate = {2022-06-16}, year = {2021}}

\bibitem{ToIP}
ToIP: {Achieving trusted digital transactions across the globe: OIX and ToIP align to make it happen} (2022), \url{https://trustoverip.org/news/2022/06/15/ achieving-trusted-digital-transactions-across-the-globe-oix-and-toip\\ -align-to-make-it-happen/}

\bibitem{tsakalakis2017identity}
Tsakalakis, N., Stalla-Bourdillon, S., O'Hara, K.: Identity assurance in the uk: technical implementation and legal implications under eidas. The Journal of Web Science  \textbf{3}(3),  32--46 (2017)

\bibitem{turkanovic2020signing}
Turkanovi{\'c}, M., Podgorelec, B.: Signing blockchain transactions using qualified certificates. IEEE Internet Computing  \textbf{24}(6),  37--43 (2020)

\bibitem{vasileedis}
VASILE, P.C., DINU, A.: edis--electronic diploma integrity service

\bibitem{voigt2017eu}
Voigt, P., Von~dem Bussche, A.: The eu general data protection regulation (gdpr). A Practical Guide, 1st Ed., Cham: Springer International Publishing  (2017)

\end{thebibliography}
\end{document}